\documentclass[conference]{IEEEtran}
\IEEEoverridecommandlockouts

\usepackage{cite}
\usepackage{amsmath,amssymb,amsfonts}
\usepackage{algorithmic}
\usepackage{graphicx}
\usepackage{braket}
\usepackage{quantikz}
\usepackage{textcomp}
\usepackage{xcolor}
\usepackage{quantikz}
\usepackage{lipsum} 
\usepackage{tabularx,booktabs}
\usepackage{multirow}
\usepackage{comment}
\usepackage{hyperref}
\usepackage{orcidlink}
\usepackage{lipsum,booktabs}
\newcolumntype{C}{>{\centering\arraybackslash}X} 
\usepackage[ruled,vlined,linesnumbered]{algorithm2e}
\def\BibTeX{{\rm B\kern-.05em{\sc i\kern-.025em b}\kern-.08em
    T\kern-.1667em\lower.7ex\hbox{E}\kern-.125emX}}

\def\ie{{\em i.e., }}
\def\eg{{\em e.g., }}
\def\reals{\mathbb{R}}
\def\iu{\mathrm{i}}
\newtheorem{definition}{Definition}
\newtheorem{theorem}{Theorem}

\title{Verification of Quantum Circuits through Barrier Certificates using a Scenario Approach
}
\author{ \IEEEauthorblockN{Siwei Hu\IEEEauthorrefmark{2} \orcidlink{0009-0009-1040-8839}, Victor Lopata\IEEEauthorrefmark{2} \orcidlink{0009-0007-3353-1325} , Sadegh Soudjani\IEEEauthorrefmark{3} \orcidlink{0000-0003-1922-6678},  Paolo Zuliani\IEEEauthorrefmark{2}\textsuperscript{,}\IEEEauthorrefmark{4} \orcidlink{0000-0001-6033-5919} \ } \IEEEauthorblockA{\IEEEauthorblockA{\IEEEauthorrefmark{2}Dipartimento di Informatica, Università di Roma “La Sapienza”, Rome, Italy\IEEEauthorblockA{\IEEEauthorrefmark{3}Max Planck Institute for Software Systems, Kaiserslautern, Germany \IEEEauthorblockA{\IEEEauthorrefmark{4} Email: \href{mailto:zuliani@di.uniroma1.it}{zuliani@di.uniroma1.it}} }}}}

\begin{document}

\maketitle

\begin{abstract}
In recent years, various techniques have been explored for the verification of quantum circuits, including the use of barrier certificates, mathematical tools capable of demonstrating the correctness of such systems. These certificates ensure that, starting from initial states and applying the system's dynamics, the system will never reach undesired states. In this paper, we propose a methodology for synthesizing such certificates for quantum circuits using a scenario-based approach, for both finite and infinite time horizons. In addition, our approach can handle uncertainty in the initial states and in the system's dynamics. We present several case studies on quantum circuits, comparing the performance of different types of barrier certificate and analyzing which one is most suitable for each case. 
\end{abstract}


\section{Introduction}\label{sec:intro}

In the past few months, there has been tremendous progress towards fault-tolerant quantum computing, \ie the ability to run arbitrarily long quantum computations with (small) bounded error. For example, both Google \cite{Willow} and Amazon \cite{Ocelot} have significantly refined their quantum chips and experimentally validated the efficacy of quantum error correction codes. Again, Microsoft have experimentally demonstrated for the first time a topological quantum computing \cite{Kitaev-FTQC} approach, which would lead to inherently more robust qubits and gates by their physical nature \cite{Majorana1}. While the works above are merely proof-of-concepts involving single qubits, they start the move away from NISQ \cite{PreskillNISQ} systems and bring us considerably closer to the goal of fault-tolerance quantum computing with the so-called Megaquop machines\cite{Preskill2025}. In such a rapidly evolving landscape, it is even more important to ensure that quantum computers behave as expected. Since quantum circuits are by far the most used means of implementing quantum algorithms, the problem then becomes the verification of quantum circuits.

Unfortunately, verifying quantum systems is hard. It is well known that simulating general quantum systems is considered a difficult problem for which there is currently no efficient (classical) solution \cite{Feynman1982}. Hence, unlike for classical circuits, testing cannot be a sound and feasible approach to verifying general (non-stabilizer \cite{StabSim}) quantum circuits. Therefore, different approaches must be considered. 

In this paper, we propose an efficient method for generating certificates that formally prove the correctness of a quantum circuit with respect to a set of initial states and a set of undesired (or unsafe) states. If our method finds such a certificate, then we are guaranteed that for no one of the specified initial states the circuit will eventually reach an undesired state (this is also known as the {\em reachability problem}). We note that both the initial set and the undesired set may be continuous, and the time horizon of the circuit's evolution may be unbounded. 

Our approach is based on {\em barrier certificates} \cite{prajna2004bc}, which are essentially real functions of the state space of the system that act as ``barriers'' between the reachable states of the system and the undesired states.  Checking whether a function is an actual barrier certificate can be done automatically by SMT (Satisfiability Modulo Theory) solvers. However, finding candidate barrier certificates is difficult, as the reachability problem is in general undecidable \cite{AlurHA}. To combat this issue, we extend to quantum circuits a scenario-based approach (\ie based on sampling) for synthesizing barrier certificates for real systems \cite{scenario-synth}. We apply our technique to synthesize barrier certificates for quantum circuits over both bounded and unbounded time horizons, and for quantum circuits with uncertain behavior (\eg for Grover's quantum search algorithm \cite{Grover96}).

\section{Related Work}
The literature on quantum computing verification is growing fast, so here we focus on quantum circuits only. Many different formal techniques from classical verification have been adapted to the quantum context. An early work verifies quantum circuits by transforming them into quantum Markov chains, which are then model checked \cite{Anticoli16}. The authors of \cite{QBricks} develop a deductive approach to quantum circuit verification using SMT solvers. Again, the Giallar tool \cite{Giallar} exploits SMT solvers to verify quantum circuit compiler passes (\ie that the quantum semantics is preserved at each pass). Abstract interpretation is another classical verification method that has been extended to quantum circuits \cite{QuantumAbstractInterp}. On a similar vein, a static analyzer for finding families of bugs in Qiskit quantum circuits is presented in \cite{LintQ}. Two recent works develop instead an automata-based approach \cite{Chen23a} and the AutoQ tool \cite{Chen23b} that can verify relational properties of quantum circuits.

The use of barrier certificates for the verification of quantum circuits was introduced in \cite{lewis2024barrier}, but the approach is restricted to polynomial barrier functions while our technique does not have this constraint. In addition, a common shortcoming of previous works is their inability of handling uncertain quantum circuits, \ie circuits that may deviate from their nominal behavior.  Our approach does not suffer from this limitation.

For more on the formal verification of quantum systems, including quantum programs, we refer the reader to two recent surveys \cite{lewis2022formal,FMQCSurvey} with tools available for automating the verification task (e.g., \cite{lewis2024automated}). 

\section{Background on Barrier Certificates}
In this section, we cover the fundamentals of barrier certificates for discrete-time real systems under uncertainty.
\begin{definition}[Discrete-time real-space system]\label{def: discreterealssys}
    A discrete-time dynamical system $\mathcal{G}$ in $\mathbb{R}^n$, for some $n \in \mathbb{N}$, is denoted by a state set $\mathcal{X} \subseteq \mathbb{R}^n$ and a set-valued transition map $f:\mathcal{X}\rightrightarrows   \mathcal{X}$. The system evolves according to 
    \begin{equation*}
        \mathcal{G}: x_{t+1} \in f(x_t).
    \end{equation*}
    where $x_t$ represents the state of the system at time $t \in \mathbb{N}$.
\end{definition}
Note that the set-valued transition map $f$ captures the effect of uncertainty in the system. A deterministic system will have $f(x)$ a singleton set for all $x\in \mathcal{X}$.
The sequence $\langle x_0, x_1, \hdots \rangle$, such that $x_{i+1} \in f(x_i)$ for all $i \in \mathbb{N}$, is defined as \textit{trace} or \textit{state sequence}. 

\subsection{Invariant Barrier Certificate}
\begin{definition}
    A system $\mathcal{G}: x_{t+1} \in f(x_t)$, evolving over $\mathcal{X} \subseteq R^n$, is (invariantly) safe if, starting from the initial state set $\mathcal{X}_0 \subseteq \mathcal{X}$, it will never reach the unsafe state set $\mathcal{X}_u \subseteq X$. That is for any trace $\langle x_0, x_1, \hdots \rangle$, such that $x_0 \in \mathcal{X}_0$, then $x_i \not \in \mathcal{X}_u$ for all $i \in \mathbb{N}$.
\end{definition}

The \textit{safety verification problem} is to determine wether a given system is safe with respect to $\mathcal{X}_0$ and $\mathcal{X}_u$. There are several approaches to system safety verification, one of which is the use of \textit{barrier certificates} \cite{prajna2004bc}. These are real functions of the state space that allow us to prove that no system trajectory can reach undesirable states.
\begin{definition}\cite{prajna2004bc}\label{def: bc1}
    We define a \textit{barrier certificate} for a dynamical system $\mathcal{G}$, with respect to a set of initial states $\mathcal{X}_0 \subseteq \mathcal{X}$ and a set of unsafe states $\mathcal{X}_u \subseteq \mathcal{X}$, as a function $B: \mathcal{X} \to \mathbb{R}$ that satisfies the following constraints:
    \begin{align}
    B(x) &\leq 0, \quad \forall x \in \mathcal{X}_0; \tag{1a} \label{eq: 1a} \\
    B(x) &> 0, \quad \forall x \in \mathcal{X}_u; \tag{1b} \\
    B(x') - B(x) &\leq 0, \quad \forall x \in \mathcal{X} \text{ and } \forall x'\in f(x). \tag{1c} \label{eq: 1c}
\end{align}
\end{definition}
These conditions ensure that the sequence of states, starting from the initial states \( \mathcal{X}_0 \), will never reach the undesirable region \( \mathcal{X}_u \) for any time step \( t \geq 0 \). Indeed, the last condition, also called the \textit{decrement constraint}, ensures that the barrier function does not increase along the system trajectories and the value of the certificate will always be non-increasing. Intuitively, these conditions mean that the zero level-set of $B(x)$ acts as a “barrier” between initial and unsafe regions. All initial states lie in the region $B(x)\le 0$, while all unsafe states lie in $B(x)>0$, and the system’s dynamics cannot drive the barrier function upwards. The existence of this type of certificate formally verifies that the system is safe. For detailed proofs, we refer the interested readers to \cite{prajna2004bc}. Condition \eqref{eq: 1c} is often too strict to satisfy, for this reason, it is common to relax the certificate constraints (while preserving its correctness), thereby expanding the set of functions that can be used to verify the system. Therefore, we introduce the \textit{k}-inductive barrier certificates.
\begin{definition}\cite{safkind}\label{def:k-ind}
    We define a \textit{k-inductive barrier certificate} for a dynamical system $\mathcal{G}$, with respect to a set of initial states $\mathcal{X}_0 \subseteq \mathcal{X}$ and a set of unsafe states $\mathcal{X}_u \subseteq \mathcal{X}$, as a function $B: \mathcal{X} \to \mathbb{R}$ if there exist $k \in \mathbb{N}_{\geq 1}, \epsilon \in \mathbb{R}_{\geq 0}$, and $d > k \epsilon$ such that the following conditions hold:
    \begin{align}
        B(x) &\leq 0, \quad \forall x \in \mathcal{X}_0; \tag{2a} \label{eq:2a} \\
        B(x) &\geq d, \quad \forall x \in \mathcal{X}_u; \tag{2b} \label{eq:2b}\\
        B(x') - B(x) &\leq \epsilon, \quad \forall x \in \mathcal{X} \text{ and } \forall x'\in f(x); \tag{2c} \label{eq:2c} \\
        B(x'') - B(x) &\leq 0 , \quad \forall x\in \mathcal{X} \text{ and } \forall x''\in f^k(x). \tag{2d} \label{eq:2d}
    \end{align}
\end{definition}

In the above definition, the restrictive monotonicity condition \eqref{eq: 1c} is relaxed by allowing $B$ to increase slightly in the next transitions \eqref{eq:2c}, as long as it does not increase overall over any $k$-step window \eqref{eq:2d}. The condition $d > k \epsilon$ guarantees that $B$ cannot drift from the safe region to the unsafe region, even if it increments by $\epsilon$ for $k-1$ consecutive steps. The existence of this type of certificate formally verifies that the system is safe. For detailed proofs and better understanding the certificate's behavior, we refer the interested readers to \cite{safkind}.

\subsection{Finite Horizon Barrier Certificate}

When system safety is only required over a fixed time interval, we can relax the invariant conditions to obtain a \emph{finite horizon barrier certificate}. This is particularly useful when verifying the safety of quantum circuits that operate over a finite number of steps (for example in the Grover's algorithm \cite{Grover96}).

\begin{definition}[Finite Horizon Barrier Certificate\cite{BCfinHorizonSadegh}]\label{def: finiteHor}
Let $\mathcal{G}: x_{t+1} \in f(x_t)$ be a discrete-time system with state space $\mathcal{X} \subseteq \mathbb{R}^n$, the set of initial states $\mathcal{X}_0 \subseteq \mathcal{X}$, and the set of unsafe states $\mathcal{X}_u \subseteq \mathcal{X}$. Given a time horizon $T\in \mathbb{N}$, a function $B:\mathcal{X}\to\mathbb{R}$ is called a \emph{finite horizon barrier certificate} (or barrier certificate over a time horizon $[0,T]$) if there exists $\delta \in \mathbb{R}_{\geq 0}$ and $\gamma, \lambda \in \mathbb{R}$, such that $\gamma + \delta \cdot T < \lambda$, and
\begin{align}
B(x) &\leq \gamma, \quad \forall x \in \mathcal{X}_0; \tag{5a}\\
B(x) &\geq \lambda, \quad \forall x \in \mathcal{X}_u; \tag{5b}\\
B(x') - B(x) &\leq \delta, \quad \forall x \in \mathcal{X}
\text{ and } \forall x'\in f(x).\tag{5c}
\end{align}
\end{definition} 
\begin{theorem} 
    Let $\mathcal{G}:x_{t+1} \in f(x_t)$ be a discrete-time dynamical system. If there exists a barrier certificate $B:\mathcal{X}\rightarrow\mathbb{R}$ according to Definition \ref{def: finiteHor}, with respect to initial set $\mathcal{X}_0 \subseteq \mathcal{X}$ and unsafe set $\mathcal{X}_u \subseteq \mathcal{X}$, then any sequence of states $\langle x_0, \hdots, x_T\rangle$, with $x_0 \in \mathcal{X}_0$, satisfies $x_i \not \in \mathcal{X}_u$ for all $0\leq i \leq T$.
\end{theorem}

In essence, the finite horizon certificate allows the barrier to grow slightly at each step, provided  that within the time window $[0,T]$, the evolution of the system does not lead to any unsafe state. For a detailed proof, see \cite{BCfinHorizonSadegh}.

\section{Problem Statement}
The goal of our work is to prove the correctness of quantum circuit properties with respect to a set of initial states and a set of unsafe states using barrier certificates. For example, we would like to show that the repeated application of the Z gate does not change the probability of measuring either of the basis state of a qubit (recall that the Z gate operates a `phase-flip' only). The first step is to establish a connection between the concept of barrier certificates and quantum circuits.

\subsection{Quantum Circuits as Dynamical System}
Quantum circuits differ from classical dynamical systems in that their states belong to a complex Hilbert space and evolve under unitary operations. Our goal is therefore to model the quantum circuit as a complex discrete-time system.
\begin{definition}[Discrete-time complex-space system \cite{lewis2024barrier}]\label{def: discrete time complex}
    A discrete-time complex-space system is a tuple $S = (\mathcal{Z},\mathcal{Z}_0, F, f)$, where:
    \begin{itemize}
        \item $\mathcal{Z} \subseteq \mathbb{C}^n$ is the continuous complex-valued state space;
        \item $\mathcal{Z}_0 \subseteq \mathcal{Z}$ is the set of initial states;
        \item $F$ is a finite set of set-valued functions that contains all possible dynamics the system can perform and
        \item $f:\mathbb{Z}_{\geq0} \rightarrow F$ is a function that assigns at each time step the dynamics of the system.
    \end{itemize}
    The system is therefore defined by the following dynamics:
    \begin{equation*}
        z_{t+1} \in f(t)(z_t) = f_tz_t .
    \end{equation*}
\end{definition}

Note that this model does not capture conditional selection of the transition function, which we leave for future work.

\begin{figure}[h]
\centering
    \scalebox{0.9}{
    \begin{quantikz}
    \lstick{$\ket{0}$}&\gate{Z}\gategroup[3, steps=2]{$U_0$}&\gate{H}&&\targ{}\gategroup[3, steps=2]{$U_1$}&\ctrl[open]{1}&&\gate{Z}\gategroup[3, steps=2]{$U_0$}&\gate{H}& \text{ } \cdots \\
    \lstick{$\ket{0}$}&\gate{Z}&\gate{H}&&\ctrl[open]{-1}&\targ{}&&\gate{Z}&\gate{H}& \text{ } \cdots \\
    \lstick{$\ket{0}$}&\gate{Z}&\gate{H}&&\gate{Y}&\gate{H}&&\gate{Z}&\gate{H}& \text{ }\cdots 
    \end{quantikz}
    }
\caption{An infinite-depth quantum circuit alternating unitaries $U_0$ and $U_1$.}
\label{fig: DiscreteTime Circuit}
\end{figure}
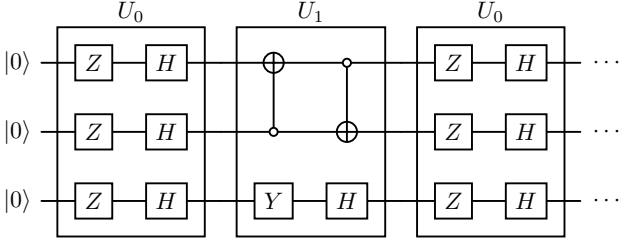
In this new definition of discrete-time system, the set-valued transition function is not fixed (as was for Definition~\ref{def: discreterealssys}); instead, it is chosen from a finite set $F$, meaning that the evolution can change over time. This makes the system more versatile and suitable for modeling a quantum circuit, where its evolution changes based on the quantum gates. Note also that we can use Definition \ref{def: discreterealssys} to model our quantum circuit, adapting it to the complex space. However, this modification makes the evolution of the circuit less flexible. Later, we will examine examples that will allow us to observe the behaviors of the barrier certificates on these two types of systems.

We model a quantum circuit as a discrete-time complex-space quantum dynamical system
\[
\ket{\psi_{k+1}} \in U_k \ket{\psi_k},
\]
where $\ket{\psi_k}\in \mathcal{Z}\subseteq \mathbb{C}^{2^n}$ is the state at step $k$, and $\{U_k\}_{k\ge0}$ is the sequence of (possibly uncertain) unitary operators defining the circuit dynamics.
As an example, consider the quantum circuit shown in Figure \ref{fig: DiscreteTime Circuit}; describing it as a system of Definition \ref{def: discrete time complex} is straightforward:
\begin{itemize}
    \item $\mathcal{Z} = \{z \in \mathbb{C}^8:\sum_j |z_j|^2 = 1\}$;
    \item $\mathcal{Z}_0 = \{z \in \mathcal{Z} : |z_0|^2 \geq 0.99\}$;
    \item $F = \{U_0, U_1\}$ and
    \item $f_t(z) =     \begin{cases}
                            U_0 z & \textbf{if } t = 0,2,4,\ldots \\
                            U_1 z & \textbf{if } t = 1,3,5,\ldots;
                        \end{cases}$
\end{itemize}
Consequently, the tuple $S = (\mathcal{Z}, \mathcal{Z}_0, F, f)$ is the dynamic system that models the quantum circuit in Figure \ref{fig: DiscreteTime Circuit}. In this example, the uncertain dynamics might be represented by an uncertain version of the Hadamard gate $H$ in the unitary $U_0$. An uncertain Hadamard could be defined by the unitary operator $
    H_\epsilon = \frac{1}{\sqrt{1+\epsilon^2}} \begin{psmallmatrix}
        1 & \epsilon \\
        \epsilon & -1
    \end{psmallmatrix}
$, for any $\epsilon$ ``close'' to 1.

\subsection{Barrier Certificates for Quantum Circuits}
\begin{definition}[Safety for Complex Systems]
    Let $S = (\mathcal{Z}, \mathcal{Z}_0, F, f)$ be a complex discrete-time system, and let $\mathcal{Z}_u$ be the set of unsafe states. The system is considered safe with respect to $\mathcal{Z}_u$ if, for every initial state $z_0 \in \mathcal{Z}_0$, the state $z_t \not \in \mathcal{Z}_u$ $\forall t\in\mathbb{Z}_{\geq 0}$.
\end{definition}

The model dynamics introduced by Definition \ref{def: discrete time complex} is time-varying, and we thus enlarge our class of certificates by considering time-varying barrier certificates that are less restrictive.

\begin{theorem}[k-Inductive Hybrid Barrier Certificate~\cite{lewis2024barrier}]\label{thm:k-inductive}
Consider a discrete-time complex-space system 
\[
\mathcal{G} = (\mathcal{Z}, \mathcal{Z}_0, F, f),
\]
where $\mathcal{Z}\subseteq \mathbb{C}^{2^n}$ is the state space, $\mathcal{Z}_0$ is the initial set, and $f_t:\mathcal{X}\rightrightarrows\mathcal{X}$ is the time-dependent transition function with $f_t\in F$. Consider also an unsafe state set $\mathcal{Z}_u$. Suppose there exists a family of functions $\{B_t(z)\}_{t\ge0}$ with $B_t:\mathcal{Z}\to\mathbb{R}$ and constants $k\ge1$, $\epsilon,\gamma\ge0$, and $d> k(\epsilon+\gamma)$ satisfying
\begin{align}
&B_0(z) \leq 0,\ \forall z\in\mathcal{Z}_0; \tag{3a}\\
&B_t(z) \geq d,\ \forall z\in\mathcal{Z}_u,\ \forall t\ge0; \tag{3b}\\
&B_t(z') - B_t(z) \leq \epsilon,\ \forall z\in\mathcal{Z},\ \forall z'\in f_t(z),\ \forall t\ge0; \tag{3c}\\
&B_{t+1}(z) - B_t(z) \leq \gamma,\ \forall z\in\mathcal{Z},\ \forall t\ge0; \tag{3d}\\
&B_{t+k}(z'') - B_t(z) \leq 0,\ \forall z\in\mathcal{Z},\notag \\ &\quad\quad\quad\forall z''\in f_{t+k-1}\circ \cdots \circ f_t(z),\ \forall t= r k,\ r\in\mathbb{N}_{\ge0}, \tag{3e}
\end{align}
then, the system $S$ is safe with respect to $\mathcal{Z}_u$.
\end{theorem}

This k-inductive hybrid approach allows the barrier certificate to vary over time (thus accommodating temporary increases) provided that over every $k$-step interval the overall change is non-positive.

However, constructing such barrier functions analytically for a given quantum circuit is challenging. One of the challenges is
the return value of the certificate itself. In our case, the barrier certificate's domain is the space of quantum states, which are vectors of complex numbers with Euclidean norm equal to 1; however, the return value cannot belong to the set of complex numbers, as they do not admit a total order compatible with algebraic operations \cite{lewis2024barrier}. 
In the next section, we introduce the computational approach used to synthesize barrier certificates and show how the constraints are handled.

\section{Scenario-based Barrier Certificates for Quantum Circuits}\label{Section: scenariobased}
In this section, we present our approach to synthesizing barrier certificates. By considering barrier {\em templates}, \ie parametrized families of candidate functions, finding barrier certificates can then be reduced to solving an optimization problem. We define a {\em template} as a family of functions $\{ B(\alpha, z): \mathcal{Z} \rightarrow \reals  \mid \alpha \in A \}$, where $A$ is the parameter domain, and by fixing the parameter vector $\alpha$ one obtains a barrier certificate candidate. Commonly used templates include polynomials and neural networks \cite{Fossil2}, which clearly can be framed in the definition just given. Now, the constraints that define a barrier certificate can be mapped to a constrained optimization problem over the parameter domain $A$, through which the parameters of the candidate certificate are determined. For example, to synthesize the barrier certificate of Definition \ref{def: bc1}, we solve the constrained optimization problem:
\begin{equation}
\label{eq:BCoptprob}
\begin{aligned}
    &\max_{\alpha \in A} \; \gamma \\
    &\begin{cases}
        B(\alpha, z) \leq 0, & \forall z \in \mathcal{Z}_0, \\
        B(\alpha, z) \geq \gamma, & \forall z \in \mathcal{Z}_u, \\
        B(\alpha,z') - B(\alpha,z) \leq 0 & \forall z \in \mathcal{Z},\ \forall z'\in f(z).
    \end{cases}
\end{aligned}
\end{equation}
If max \(\gamma = \gamma^* > 0\), then we have successfully found a certificate, \ie a function $B(\alpha^*, z)$ where $\alpha^*$ is a parameter vector that maximizes $\gamma$. If $\gamma^* \leq 0$, then it does not necessarily mean that our system is unsafe, but only that there is no barrier certificate within the chosen template. Then we may try different templates or weaker definitions of barrier certificate, \eg $k$-inductive.
(Note that the existence of a barrier certificate is a {\em sufficient} condition for safety, and for nearly all real dynamical systems it is also a {\em necessary} condition \cite{BCnecessary}. For quantum systems the necessity of barrier certificates is currently an open problem.)

In general, solving the optimization problem \eqref{eq:BCoptprob} is challenging, as the template usually includes nonlinear functions.

\subsection{Scenario-based Optimization}
Given a parameter domain $A$, system state space (or scenarios set) $\mathcal{Z}$, and constraint function $g:A \times \mathcal{Z} \rightarrow \reals$, there are three main approaches to optimize an objective function $\gamma : A \rightarrow \reals$, \ie finding $\alpha \in A$ that satisfies the constraints $g(\alpha,z) \leq 0$ for any $z\in \mathcal{Z}$ while minimizing the objective function $\gamma$ \cite{Garatti2024}.
\begin{itemize}
    \item \textbf{Robust Optimization}: This approach addresses the problem in the worst-case scenario; every constraint must be satisfied for all possible scenarios:
    \[
    \min_{\alpha \in A} \gamma(\alpha) \quad \text{s.t.} \quad g(\alpha, z) \leq \zeta, \; \forall z \in \mathcal{Z},
    \]
    where $\zeta$ is a tolerance threshold. This approach provides a consistently reliable solution, but at the expense of scalability when non-linear functions are involved \cite{deltadecide}.
    \item \textbf{Optimization with Probabilistic Constraints}: This approach ensures that the constraints are satisfied with a certain probability:
    \[
    \min_{\alpha \in A} \gamma(\alpha) \quad \text{s.t.} \quad \mathbb{P}[g(\alpha, z) \leq 0] \geq 1 - \epsilon
    \]
    where $\epsilon >0$ is small. This approach is more efficient than robust optimization, but the choice of a suitable probability measure $\mathbb{P}$ over the scenarios is critical for the problem's context.
    \item \textbf{Scenario-Based Optimization}: This approach relies on sampling points (or scenarios) from $\mathcal{Z}$ and ensures that the constraints must hold for these specific scenarios:
    \[
    \min_{\alpha \in A} \gamma(\alpha) \quad \text{s.t.} \quad g(\alpha, z_i) \leq 0\;\; \text{for}\;\; z_1, \hdots, z_N  \in \mathcal{Z} .
    \]
\end{itemize}

Ideally, we would like to solve our optimization problem \eqref{eq:BCoptprob} through robust optimization. This is generally infeasible, because of its complexity when nonlinear constraints and/or objectives are involved. Consequently, we adopt a scenario-based approach, which allows for a much more computationally efficient solution. The scenario-based formulation replaces the universal conditions on a barrier certificate candidate $B(\alpha,z)$ with constraints enforced only on a finite set of sample points $z_i \in \mathcal{Z}$. For example, given a deterministic quantum system $f$, the universally-quantified condition 
\[
    B(\alpha, f(z)) - B(\alpha,z) \leq 0  \quad\quad      \forall z \in \mathcal{Z}
\]
becomes a conjunction of $n$ unquantified constraints
\[
    B(\alpha, f(z_i)) - B(\alpha,z_i) \leq 0 \quad\quad       z_1, \ldots, z_n \in  \mathcal{Z},
\]
where $n$ is the number of sampled scenarios. Our original optimization problem \eqref{eq:BCoptprob} becomes then
\begin{equation}\label{eq:sb-BCoptprob}
\begin{aligned}
    &\max_{\alpha \in A} \; \gamma \\
    &\begin{cases}
        B(\alpha, z_i) \leq 0, & z_i \in \mathcal{Z}_0, \\
        B(\alpha, z_i) \geq \gamma, & z_i \in \mathcal{Z}_u, \\
        B(\alpha,f(z_i)) - B(\alpha,z_i) \leq 0 & z_i \in \mathcal{Z},
    \end{cases}
\end{aligned}
\end{equation}
where for $i = 1, \ldots, N$ each $z_i$ is a point sampled in the sets $\mathcal{Z}_0, \mathcal{Z}_U,$ and $\mathcal{Z}$, depending on the constraint (\ie the total number of constraints is $3N$). For uncertain quantum systems, the space of uncertainty is also sampled for each $z_i$.

The scenario-based approach has three main benefits:
\begin{enumerate}
    \item If the template functions are linear in the parameter $\alpha$, then problem \eqref{eq:sb-BCoptprob} reduces to a {\em linear programming} problem, for which efficient algorithms exists (\eg simplex and interior-point algorithms). This greatly reduces the complexity of barrier certificate candidate synthesis. 
    \item If problem \eqref{eq:sb-BCoptprob} returns a $\gamma^* \leq 0$, then we know for sure that the same applies to problem \eqref{eq:BCoptprob}. As explained earlier in this section, we cannot conclude that the system is unsafe, but we may need to change template or use a weaker notion of barrier certificate. 
    \item A candidate barrier certificate that satisfies \eqref{eq:sb-BCoptprob} might just as well satisfy the stricter problem \eqref{eq:BCoptprob}, thereby formally showing that our system is safe (over the entire space, not just for the sampled scenarios). As mentioned in Section \ref{sec:intro}, checking the universally-quantified constraints of \eqref{eq:BCoptprob} for a candidate barrier certificate can be done rather efficiently using SMT solvers. If the candidate function does not satisfy some of the constraints, then we may sample more scenarios, solve problem \eqref{eq:sb-BCoptprob} and SMT-check again, or change template functions, or simply stop altogether.
\end{enumerate}

Two disadvantages of the scenario-based approach are, first, the choice of the barrier template is paramount: it should be able to capture a large family of candidate barrier functions with a minimal number of parameters, to reduce the size of the linear programming problem resulting from \eqref{eq:sb-BCoptprob}. Second, while SMT solvers have shown to be empirically efficient, the computational complexity of the problem they solve is exponential in the number of variables (in our case, the number of qubits in the system).

\subsection{Constraint Formulation for Polynomial Templates}
For simplicity we present our algorithm for deterministic quantum systems and using polynomial templates, although other templates may be easily accommodated. While formulating the constraints for the optimization problem \eqref{eq:sb-BCoptprob}, it is important to consider how quantum states are represented. Since the probability amplitudes of quantum states are expressed as complex numbers, we assume that the scenarios $z_i$'s are complex vectors, sampled from a specific region (\eg  the initial set of states $\mathcal{Z}_0$). We choose a degree and we define the (polynomial) barrier certificate template as
\begin{equation*}
    B(\alpha,z) = \sum_i \alpha_i  t_i (z)
\end{equation*}
where
\begin{itemize}
    \item the $\alpha_i$'s are the polynomial's unknown complex coefficients to be determined via the linear program \eqref{eq:sb-BCoptprob}; and 
    \item the $t_i(z)$'s are the terms of the polynomial with complex variables,
\end{itemize}
and the sum is over all the possible terms appearing in a polynomial with the given degree.

After sampling the scenarios, the terms $t_i(z)$'s will become complex numbers, which we denote as $c_i = t_i(z)$. The $i$-th term of our candidate polynomial certificate thus has the form $\alpha_i \times c_i$. We have that
\begin{equation*}
\begin{array}{l}
    \alpha_i \times c_i = (\text{Re}(\alpha_i) + \iu \text{Im}(\alpha_i))\times(\text{Re}(c_i) + \iu \text{Im}(c_i))  \\ \\
     = \text{Re}(\alpha_i)\text{Re}(c_i) - \text{Im}(\alpha_i)\text{Im}(c_i) + \iu(\text{Re}(\alpha_i)\text{Im}(c_i) + \\\text{Re}(c_i)\text{Im}(\alpha_i)) .
\end{array}
\end{equation*}
Since we want the certificate to produce a real value, we eliminate the imaginary part and work only with the real part.
Consequently, our polynomial will take the following form:
\begin{equation*}
    B(a, z) = \sum_i \text{Re}(\alpha_i)\text{Re}(c_i) - \text{Im}(\alpha_i)\text{Im}(c_i),
\end{equation*}
where we note again that
\begin{itemize}
    \item $\text{Re}(c_i)$ and $\text{Im}(c_i)$ are the values computed from the sampled scenarios; and
    \item $\text{Re}(\alpha_i)$ and $\text{Im}(\alpha_i)$ are the (unknown) coefficients of the polynomial certificate.
\end{itemize}

In this way, we have successfully obtained a real value from a polynomial with complex variables and coefficients, which is exactly the type of certificate we aim to find. Using this approach, we proceed to construct the three types of constraints that the polynomial barrier certificate must satisfy:
\begin{itemize}
    \item \textbf{Constraints on the initial states:}
    \begin{equation*}
        \sum_i \text{Re}(\alpha_i)\text{Re}(c_i) - \text{Im}(\alpha_i)\text{Im}(c_i) \leq 0
    \end{equation*}
    where the $c_i$'s are computed from the scenarios sampled from $\mathcal{Z}_0$.
    \item \textbf{Constraints on the unsafe states:}
    \begin{equation*}
        \sum_i \text{Re}(\alpha_i)\text{Re}(c_i) - \text{Im}(\alpha_i)\text{Im}(c_i) \geq \gamma
    \end{equation*}
    where the $c_i$'s are computed from the scenarios sampled from $\mathcal{Z}_u$.
    \item \textbf{Constraints on the system dynamics:}
    \begin{equation*}
    \begin{array}{l}
         \Big ( \sum_i \text{Re}(\alpha_i)\text{Re}(f(c_i)) - \text{Im}(\alpha_i)\text{Im}(f(c_i)) \Big ) -  \\
         \Big ( \sum_i \text{Re}(\alpha_i)\text{Re}(c_i) - \text{Im}(\alpha_i)\text{Im}(c_i) \Big ) \leq 0
    \end{array}
    \end{equation*}
    where $f$ is the system dynamics (\ie a quantum gate), and the $c_i$'s are computed from the scenarios sampled from $\mathcal{Z}$.
\end{itemize}
We recall that if we sample $N$ scenarios in each of $\mathcal{Z}_0, \mathcal{Z}_u$, and $\mathcal{Z}$, then the linear program \eqref{eq:sb-BCoptprob} will have $3N$ constraints.

\subsection{Barrier Certificates Synthesis}
We now present our algorithm for the synthesis of barrier certificates for discrete-time quantum dynamical systems, given in Algorithm \ref{algo}. The algorithm aims to find a barrier certificate in order to verify the correctness of the system with respect to a set of initial states $\mathcal{Z}_0$  and a set of unsafe states $\mathcal{Z}_u$. We now analyze the key steps of the algorithm:

\begin{algorithm}
    \caption{Barrier certificate synthesis}
    \label{algo}
    \KwIn{quantum circuit, \texttt{n\_samples}, $\mathcal{Z}_0, \mathcal{Z}_u, \mathcal{Z}$, \texttt{deg};}
    \KwOut{barrier certificate or `not found';}
    \textbf{Generate Terms}: all possible terms of max degree \texttt{deg} are generated\;\label{alg:genterms}
    \textbf{Set:}  initialize template for  barrier certificate $B(\alpha,z)$\;
    \For{t in terms}{
        \textbf{Add term $t$ to the polynomial} $B(\alpha,z)$\;\label{alg:addterm}
        \textbf{Sample scenarios from $\mathcal{Z}, \mathcal{Z}_0,$ and $\mathcal{Z}_u$:} For each set, sample \texttt{n\_samples} scenarios (or points)\;\label{alg:sampling}
        \textbf{Generate BC constraints and LP:} Generation of the barrier certificate constraints and of the linear programming (LP) problem \eqref{eq:sb-BCoptprob}\;\label{alg:genBC}
        \textbf{Solve LP:} The LP problem is solved using an appropriate linear solver\;\label{alg:solveLP}
        \If{solver returns `{\em optimum} $ > 0$'}{
        \textbf{Candidate BC:} Build the candidate certificate $B(\alpha^*,z)$ using the coefficients $\alpha^*$ obtained by solving the LP problem\;\label{alg:candidateBC}
        \textbf{Verification:} Check the candidate $B(\alpha^*, z)$ using an appropriate SMT solver\;\label{alg:verify}
        \If{solver returns `unsat'}{
        \textbf{return} $B(\alpha^*, z)$\;
        }
        }
    }
    \textbf{return} `Not found'
\end{algorithm}
{\em Line \ref{alg:genterms}-\ref{alg:addterm} (Terms Generation and Template Construction):}
In this phase, all possible monomials of degree \texttt{deg} or lower are generated. The terms will be used incrementally to construct a candidate barrier certificate $B(\alpha,z)$: each iteration of the {\bf for} loop adds a new term to the candidate.

{\em Line \ref{alg:sampling} (Sampling):}
In this part, we need to sample, say $N$, scenarios from each of the regions $\mathcal{Z}_0, \mathcal{Z}_u$, and $\mathcal{Z}$. To ensure a better coverage of the regions, we employ quasi-Monte Carlo sampling via a Sobol sequence \cite{sobol1967distribution}. If the quantum circuit under study has $n$ qubits, then the circuit state expressed in the computational basis is
\[
\ket{\psi} = \sum_{j=0}^{2^n-1} z_j |j\rangle, \quad \text{with } \sum_{j=0}^{2^n-1} |z_j|^2 = 1,
\]
and we decompose it into its real and imaginary components, so that each complex amplitude $z_j = x_j + \iu y_j$ is represented by the real variables $x_j$ and $y_j$. 
Each point of the Sobol sequence is mapped onto a quantum state by first partitioning the components to yield a probability vector $(p_0, \dots, p_{2^n-1})$, satisfying $\sum_j p_j = 1$, and then assigning independent random phases $\theta_j \in [0, 2\pi)$, so that
\( z_j = \sqrt{p_j}e^{\iu\theta_j} \).
This construction ensures that every sampled state respects the normalization constraint.

{\em Line \ref{alg:genBC}-\ref{alg:solveLP} (Generate Constraints and Optimization):}
Once the scenarios are sampled, the constraints that should be satisfied by  barrier certificates are constructed, along with the linear programming (LP) problem \eqref{eq:sb-BCoptprob}. The problem is then provided to an LP solver, which checks whether the optimum is strictly positive. If that is not the case, the process is repeated by adding one more term to the polynomial. Otherwise, we have found a candidate for the barrier certificate that satisfies the constraints for the scenarios. (The number of samples is crucial for ensuring the quality of the solution and depends on the number of qubits in the quantum system. Typically, for a single qubit between 200 and 300 samples per region are sufficient.)

{\em Line \ref{alg:candidateBC} (Build Candidate Barrier Certificate):} The vector $\alpha^*$ of the coefficients that maximize the LP problem is used to build the (polynomial) barrier certificate candidate $B(\alpha^*, z)$.

{\em Line \ref{alg:verify} (Verify Candidate Barrier Certificate):}
Once a candidate certificate has been obtained, we must check whether it satisfies the (stricter) conditions of \eqref{eq:BCoptprob}: This is done by means of an SMT solver. In general, that translates to verifying the validity of logical formulas of the type $\forall z $ $\phi(z)$, where $\phi(z)$ represents a specific type of constraint. To transform this into a satisfiability problem suitable for an SMT solver, we need to negate the formula, resulting in $\exists z\, \neg \phi(z)$, which the solver can process. For example, the constraint $B(z)\leq 0, \forall z \in Z_0$ is encoded in $\exists z$ $B(z)>0$. (The other two types of barrier certificate constraint follow easily.) If $\exists z\, \neg \phi(z)$ is false ($\neg \phi(z)$ unsatisfiable), then $\phi(z)$ is true for every $z$. However, if it is true (satisfiable), then there exists a counterexample where $\phi$ is not true, meaning a state $z$ for which $\phi$ is not satisfied.


\section{Experiments}
In this section, we apply our algorithm to several quantum circuits, analyzing the obtained results, their limitations, and potential future improvements.
\subsection{Experimental Setup:}
\subsubsection{Device Details} The experiments were conducted on a MacBook Air with an M1 chip (8-core), running Sequoia version 15.3, equipped with 8 GB of RAM.

\subsubsection{Linear Programming}
To solve the linear programming problems we used the Python library \texttt{scipy.optimize}. Specifically, we used the \textit{HiGHS} \cite{huangfu2018parallelizing} solver, based on the interior point method, as it performs better than the simplex when dealing with large numbers of variables and constraints.

\subsubsection{SMT Verification}
Once the candidate certificate has been identified, we aim to formally verify that its conditions are satisfied. To do this, we use an SMT solver, following the approach described in Section \ref{Section: scenariobased}. In particular, we use \texttt{Z3} \cite{Z3}, since the certificate has a polynomial form. 
Our implementation is available at \url{https://github.com/QuantumVerification/Quantum-Verification-QSW-2025} .

\subsection{Case Studies}
The full results of our experiments are presented in Table \ref{tab: infinite horizons examples} and Table \ref{tab: finite horizons examples} for infinite and finite horizons, respectively. 
To better illustrate the synthesis of barrier certificates using our approach, we now present in more detail some case studies.
\subsubsection{Z Gate}
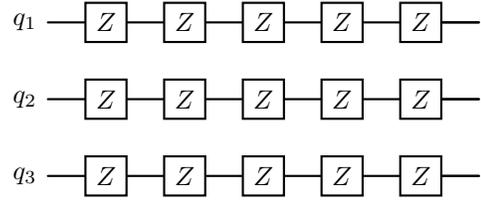
\begin{figure}
    \centering
    \begin{quantikz}
        \lstick{$q_1$} & \gate{Z} & \gate{Z} & \gate{Z} & \gate{Z} & \gate{Z} & \qw \\
        \lstick{$q_2$} & \gate{Z} & \gate{Z} & \gate{Z} & \gate{Z} & \gate{Z} & \qw \\
        \lstick{$q_3$} & \gate{Z} & \gate{Z} & \gate{Z} & \gate{Z} & \gate{Z} & \qw
    \end{quantikz}
    \caption{Quantum circuit of three qubits.}
    \label{fig: zgate}
\end{figure}

We synthesize a finite-horizon certificate for a circuit consisting of three qubits, where three Z gates are applied for five consecutive steps. Figure \ref{fig: zgate} shows the circuit, and each step of the dynamics is described by 
\begin{equation*}
    f(z) = (Z\otimes Z \otimes Z) z 
\end{equation*}
where $z$ is a three-qubit state.

Our goal is to demonstrate that, starting from initial states where the probability of measuring the state $\ket{000}$ is very high, the application of the system's dynamics for five steps will never lead to a state where the probability of measuring $\ket{001}$ exceeds 0.2. In particular, we specify as the initial and unsafe region the following sets:
\begin{equation*}
    \begin{array}{l}
         \mathcal{Z}_0 = \{z \in \mathbb{C}^8: |z_0|^2 \geq 0.9\}, \text{ and}\\
         \mathcal{Z}_u = \{z \in \mathbb{C}^8 : |z_1|^2 \geq 0.2\}. 
    \end{array}
\end{equation*}
Note that the system with respect to $\mathcal{Z}_0$ and $\mathcal{Z}_u$ is deliberately correct, as applying the Z gate will only flip the phase of the $\ket{1}$ states. By running our algorithm with an input of 20000 samples per region and a maximum degree of 2, we obtain the following barrier certificate:
\begin{equation*}
    B(z) = -9.99934z_0\overline{z_0} + 12.99994
\end{equation*}
with $\gamma = 4$, $\delta = 0$ and $\lambda = 5$. The candidate was then verified correct by the SMT solver.

\subsubsection{Alternation of CX and CZ gates}\label{example: cx cz}
\begin{figure}
    \centering
    \begin{quantikz}
        \lstick{\(|q_1\rangle\)} & \ctrl{1} & \qw      & \ctrl{1} & \qw      & \ctrl{1} & \qw \\
        \lstick{\(|q_2\rangle\)} & \targ{}  & \ctrl{-1} & \targ{}  & \ctrl{-1} & \targ{}  & \qw
    \end{quantikz}
    \caption{Quantum circuit of two qubits.}
    \label{fig: CXCZ}
\end{figure}
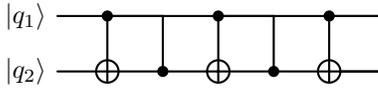

We now see how to verify a system using the (infinite-horizon) barrier certificate of Theorem \ref{thm:k-inductive}. Consider a circuit where the CX and CZ gates are applied alternately on two qubits (CX gate is applied at even time steps, and CZ otherwise, as shown in Figure \ref{fig: CXCZ}). Its dynamics is:
\begin{equation*}
    f_t(z) = 
    \begin{cases}
        \text{CX}\, z & \textbf{if } t \text{ is even} \\
        \text{CZ}\, z & \text{otherwise}
    \end{cases}
\end{equation*}
where $z$ is a two-qubit state. We specify
\begin{equation*}
    \begin{array}{l}
        \mathcal{Z}_0 = \{z \in \mathbb{C}^4: |z_0|^2 \geq 0.9\}, \text{ and}\\
         \mathcal{Z}_u = \{z \in \mathbb{C}^4: |z_1|^2 + |z_2|^2 + |z_3|^2\geq 0.5\} 
    \end{array}
\end{equation*}
as the initial and unsafe states. Running our algorithm with 20000 samples per region, a maximum degree of 2, and $k=2$, we obtain the following 2-inductive barrier certificate:
\begin{equation*}
    \begin{array}{l}
         B_0(z) = -11.7690z_0\overline{z_0} + 10.5928 \\
         B_1(z) = -11.7498z_0\overline{z_0} + 10.5828 
    \end{array}
\end{equation*}
with $\epsilon = 0.01$, $\gamma = 0.01$ and $d = 4.7079$ (see Theorem \ref{thm:k-inductive} for the meaning of these figures).

\begin{table*}[h]
  \centering
  \caption{Infinite Horizon Case Studies using Invariant Barrier Certificates (see Def. \ref{def:k-ind} and Thm. \ref{thm:k-inductive}). \underline{$\#$ Qubits} (number of qubits); \underline{Degree}(polynomial degree); \underline{$\#$ Terms} (number of polynomial terms); \underline{$\#$ Samples}(number of states sampled for each region); status( \textbf{solved}: BC found and verified, \textbf{unsolved}: no candidate BC found or candidate rejected by SMT solver, or \textbf{unknown}: candidate generated but not verified due to timeout = 300$s$). \underline{Generation Time}(average time to generate a candidate BC $\pm$ standard deviation); \underline{Verification Time}(average time to verify the candidate BC through a SMT solver $\pm$ standard deviation). Ten repetitions per experiment.} 
  \label{tab: infinite horizons examples} 
  \begin{tabular}{|c|c|c|c|c|c|c|c|@{}} 
    \toprule
    \textit{Experiment} & $\#$ Qubits & Degree & $\#$ Terms& $\#$ Samples & Status & Generation Time (s) & Verification Time (s)\\
    \midrule
    \multirow{3}{*}{Z Gate} & 3 & 2 & 3 & 15000 & \textbf{solved} & 2.45 $\pm$ 0.03 & 0.01 $\pm$ 0.01  \\
                            & 4 & 2 & 3 & 25000 & \textbf{solved}   & 6.85 $\pm$ 0.07 & 0.01 $\pm$ 0.01  \\ 
                            & 5 & 2 & 3 & 35000 & \textbf{solved}& 18.00 $\pm$ 0.45 & 0.02 $\pm$ 0.01  \\
                            \cline{2-8}
    \multirow{2}{*}{Controlled-NOT} & 2 & 2 & 3 & 2000 & \textbf{solved} & 0.23 $\pm$ 0.03 & 41.81 $\pm$ 0.03 \\
                                            
                          & 4 & 2 & 3 & 15000 & \textbf{solved} & 5.78 $\pm$ 0.06 & 35.04 $\pm$ 0.06  \\

                         \cline{2-8}
    \multirow{3}{*}{T Gate} & 3 & 2 & 3 & 15000 & \textbf{solved}  & 2.42 $\pm$ 0.03 & 0.01 $\pm$ 0.01 \\

                          & 4 & 2 & 3 & 25000 & \textbf{solved} & 6.83 $\pm$ 0.07 & 0.01 $\pm$ 0.01   \\

                          & 5 & 2 & 3 & 50000 & \textbf{solved} & 20.78 $\pm$ 1.44 &  0.02 $\pm$ 0.01 \\ \cline{2-8}

    \multirow{1}{*}{CZ Gate} & 2 & 2 & 5 & 2000 & \textbf{solved}   & 0.81 $\pm$ 0.02  & 55.28 $\pm$ 0.18  \\
                          \cline{2-8}
    \multirow{1}{*}{Hadamard} & 1 & 2 & 11 & 2000 & unknown & 5.49 $\pm$ 0.19 & TIMEOUT \\ 
                          \cline{2-8} 
    \multirow{3}{*}{X Gate} & 1 & 4 & 5 & 10000 & \textbf{solved} & 4.81 $\pm$ 0.09 & 0.21 $\pm$ 0.07  \\ 
                            & 2 & 4 & 19 & 15000 & unknown & 42.51 $\pm$ 0.65& TIMEOUT  \\
                            & 3 & 4 & 72 & 25000 & unknown & 91.98 $\pm$ 1.88 & TIMEOUT  \\
                          \cline{2-8}
    \multirow{3}{*}{SWAP Gate} & 2 & 2 & 3 & 5500 & \textbf{solved} & 7.35 $\pm$ 0.41 & 0.10 $\pm$ 0.01  \\ 
                            & 4 & 2 & 3 & 18000 & \textbf{solved} & 12.43 $\pm$ 0.15  & 0.13 $\pm$ 0.01  \\
                            & 6 & 4 & 3 & 40000 & \textbf{solved}  & 28.97 $\pm$ 0.23 & 0.6 $\pm$ 0.01 \\
                          \cline{2-8}
                          
    \multirow{2}{*}{Alternating CX and CZ} & 2 & 2 & 3 & 30000 & \textbf{solved} &  4.96 $\pm$ 0.17 &  32.53 $\pm$ 0.08 \\  
                          
                          & 4 & 2 & 3 & 40000 & \textbf{solved} & 17.15 $\pm$ 0.49  &  100.10 $\pm$ 0.01   \\ \cline{2-8}  
    \multirow{2}{*}{CX $\cdot$ CZ} & 2 & 2 & 3 & 20000 & \textbf{solved} & 16.39 $\pm$ 0.22 & 16.36 $\pm$ 0.02  \\  
                          
                          & 4 & 2 & 3 & 30000 & \textbf{solved} & 11.88 $\pm$ 0.77  &  40.04 $\pm$ 0.05 \\ \cline{2-8}   
    \multirow{2}{*}{Grover} & 2 & - & - & 4000 & unsolved & - &  - \\  
                          
                          & 5 & - & - & 30000 & unsolved & - &  - \\
    \bottomrule
  \end{tabular}
\end{table*}

\subsubsection{Grover's algorithm}\label{Grover}

The quantum search algorithm for unstructured array search provides a quadratic speed advantage compared to classical search algorithms \cite{Grover96}. Its evolution is characterized by the iterative application of the Grover operator on a quantum state initialized to the uniform superposition of all possible basis states. 
Grover's algorithm poses distinct challenges due to its inherent complexity, and initial attempts to synthesize barrier certificates using our approach failed to converge due to high-dimensional constraints (large number of terms in the candidate polynomial). This motivated a geometric reformulation of the problem.

It is well known that the evolution of Grover's algorithm can be confined to a two-dimensional subspace spanned by two orthonormal basis states $\ket{\alpha}$ and $\ket{\beta}$, representing the superposition of all non-solution states and the superposition of solution states, respectively \cite{Grover-viz}. Given an array of length $K$ whose index set is $P$ and a set $S$ of $M=|S|$ index solutions, we define 
\begin{align*}
    \ket{\alpha} &= \frac{1}{\sqrt{K-M}}\sum_{i \in P \setminus S} \ket{i}\\
    \ket{\beta} &= \frac{1}{\sqrt{M}}\sum_{i \in S} \ket{i} .
\end{align*}
The initial state $\ket{\psi_0}$ is expressed as:
\begin{align*}
      \ket{\psi_0} & = \sqrt{(K -M)/K}\ket{\alpha} + \sqrt{M/K} \ket{\beta} \\
                & =  \cos(\theta/2)\ket{\alpha} + \sin(\theta/2)\ket{\beta} .
\end{align*}
where $\theta = 2\cdot\arcsin(\sqrt{M/K})$. In the space spanned by $\ket{\alpha}$ and $\ket{\beta}$, the Grover operator $G$ is defined as:
\[
G= \begin{pmatrix}
   \cos \theta & -\sin\theta \\ 
   \sin \theta & \cos\theta  
\end{pmatrix}\] 
and therefore $G$ rotates the state by \(\theta\) radians, moving it closer to $\ket{\beta}$ with each iteration:
\[
G^k\ket{\psi_0} = \cos(\frac{2k+1}{2}\cdot\theta)\ket{\alpha} + \sin(\frac{2k+1}{2}\cdot\theta)\ket{\beta}.
\]  

\begin{figure}[ht]
    \centering
    \includegraphics[width=0.7\linewidth]{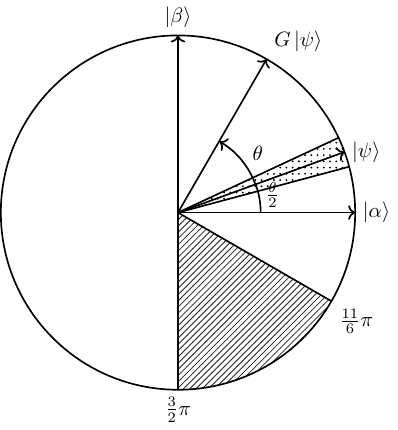}
    \caption{Representation of Grover's operator in a two-dimensional plane. The dotted region denotes $\mathcal{Z}_0$; the dashed region denotes  $\mathcal{Z}_u$.}
    \label{fig4: grover_vis}
\end{figure}

Motivated by this geometric picture, for any $\phi \in [0, 2\pi]$ we define the qubit:
\begin{equation*}
    z(\phi) = \cos \phi \ket{\alpha} + \sin \phi \ket{\beta}
\end{equation*}
where  $\phi$ is the angle between the state vector $z$ and $\ket{\alpha}$ in the two-dimensional plane (the initial state $\ket{\psi_0}$ has an angle of $\theta/2$, see Figure \ref{fig4: grover_vis}). We then define the barrier template as 
\begin{equation}\label{grover_template}
    B(c, z(\phi)) = c\cdot\phi
\end{equation}
where $c$ is a (real) coefficient to be determined.

To account for uncertainties in the initial state, we introduce a bounded perturbation \textit{err}  in $M$, defining \( M' = M \pm \textit{err} \). This perturbs the angle $\theta$ to:
\[
\theta' = 2 \cdot \arcsin\left(\sqrt{M'/{K}}\right),
\]
thereby yielding a set of initial states:
\[
\mathcal{Z}_0 = \left\{ z(\phi) \Big\vert \phi \in \left[\frac{(\theta_{min}')}{2},\frac{(\theta_{max}')}{2}\right]\right\},
\]
where $\theta_{min}' = 2 \cdot \arcsin\left(\sqrt{(M-err)/{K}}\right)$, and $\theta_{max}' = 2 \cdot \arcsin\left(\sqrt{(M+err)/{K}}\right)$.
The unsafe set \( \mathcal{Z}_u \) is defined by states with angles far from $\ket{\beta}$:  
\[
\mathcal{Z}_u = \left\{ z(\phi) \Big\vert \phi \in \left[ \frac{9}{6}\pi, \frac{11}{6}\pi\right]\right\} ,
\]
ensuring \( \mathcal{Z}_0 \) and \( \mathcal{Z}_u \) remain disjoint (see Figure \ref{fig4: grover_vis}).

\begin{table*}[h]
  \centering
  \caption{Finite Horizon Case Studies (using BC of Definition \ref{def: finiteHor} ): The fields are the same as those in Table \ref{tab: infinite horizons examples}, except for $\#$ Steps, which represents the number of steps of the quantum gate within the circuit.} 
  \label{tab: finite horizons examples}
  \begin{tabular}{|c|c|c|c|c|c|c|c|c|@{}}
    \toprule
    \textit{Experiment} & $\#$ Qubits & $\#$ Steps & $\#$ Samples & Degree & $\#$ Terms & Status & Generation Time (s) & Verification Time (s) \\
    \midrule
    \multirow{2}{*}{Z Gate} & \multirow{2}{*}{3} & 6 & 10000 & 2 & 3 & \textbf{solved} & 2.89 $\pm$ 0.05  &  0.03 $\pm$ 0.01  \\
                                                
                                              &   & 8 & 10000 & 2 & 3 & \textbf{solved} & 3.01 $\pm$ 0.04 &  0.08 $\pm$ 0.01 \\
                           \cline{2-9}
    \multirow{3}{*}{Hadamard} & \multirow{2}{*}{1} & 1 & 10000 & 2 & 3 & \textbf{solved} & 4.37 $\pm$ 0.05 & 0.02 $\pm$ 0.01  \\
                                                
                                              &   & 2 & 30000 & 2 & 11 & unknown & 43.63 $\pm$ 0.07 & TIMEOUT  \\
                                              & 2 & 1 & 40000 & - & - & unsolved & - &  - \\
                           \cline{2-9}
    \multirow{6}{*}{X Gate} & \multirow{2}{*}{1} & 1 & 500 & 4 & 5 & \textbf{solved} & 2.85 $\pm$ 0.13 & 0.09 $\pm$ 0.01  \\
                                            
                                              &  & 10 & 500 & 4 & 5 & \textbf{solved} &  2.95 $\pm$ 0.8  & 0.02 $\pm$ 0.01  \\
                          
                          & \multirow{2}{*}{2} & 1 & 10500 & 4 & 23 & unknown & 30.12 $\pm$ 0.16 & TIMEOUT  \\
                                      &  & 2 & 10500 & 4 & 23 & unknown & 25.96 $\pm$ 0.08 & TIMEOUT  \\
                          
                          & \multirow{2}{*}{3} & 1 & 18000 & 2 & 13 & \textbf{solved} & 8.48 $\pm$ 0.17 & 105.41 $\pm$ 0.04  \\ 
                          
                          &  & 2 & 18000 & 4 & 13 & unknown & 63.60 $\pm$ 0.04  & TIMEOUT  \\
                   
                          
                          
    \bottomrule
  \end{tabular}
\end{table*}

\begin{table*}[h]
  \centering
  \caption{Grover's case studies (template \ref{grover_template} in section \ref{Grover}): $\#$ Qubits, $\#$ Steps and $\#$ Samples are the same as table \ref{tab: finite horizons examples}. $M$ indicates the number of solutions, $err$ is the perturbation applied to $M$ that yields the set of initial states; $\eta$ is the perturbation applied to the rotation angle $\theta$}
  \label{tab: grover_example}
  \begin{tabular}{|c|c|c|c|c|c|c|c|}
    \hline
    $\#$ Qubits & $\#$ Steps& $M$ & $err$ & $\eta$ & $\#$ Samples & Generation Time (s) & Verification Time (s) \\
    \hline
    5 & 5 & 1 & 0.5 & 0.3 & 3000 & $ 0.49 \pm 0.02 $ & $ 0.10 \pm 0.03$ \\
    5 & 2 & 8 & 0.5 & 0.3 & 3000 & $ 0.34 \pm 0.06 $ & $ 0.08 \pm 0.03$ \\
    10 & 3 & 128 & 5 & 0.3 & 10000 & $ 1.92 \pm 0.14$ & $ 0.23\pm 0.04$ \\
    30 & 814 & 1000 & 50 & 0.003 & 20000 & $ 2.44 \pm 0.02$ & $ 0.09 \pm 0.01 $ \\
    \hline
  \end{tabular} 
\end{table*}
We define the finite horizon $T$ as the optimal number of iterations required by Grover's algorithm:
\[
T \leq \left\lceil \frac{\pi}{4} \sqrt{\frac{M}{K}} \right\rceil , 
\]
which we use to synthesize the barrier certificate of Definition \ref{def: finiteHor} for the template \eqref{grover_template}.

We now aim at taking into account uncertainty in the Grover operator $G$ itself.
Quantum systems are inherently susceptible to noise due to factors such as decoherence, gate errors or environmental interactions. These factors can perturb the ideal rotation angle $\theta$ of $G$, altering the behavior of Grover's algorithm. Under ideal conditions, Grover's algorithm gets close to $\ket{\beta}$ with $T$ iterations of $G$. But noise can disrupt this process, leading to over-rotations or under-rotations that could increase the probability of measuring non-solution states. 

We introduce a further disturbance $\mu$ to $\theta$, which leads to a noisy rotation angle $\tilde{\theta}$ in the Grover operator $G$, such that:
\[
\tilde{\theta} = \theta + \mu, \quad \mu \in [-\eta, \eta]
\]
where $\eta > 0$ bounds the maximum deviation (in absolute value) of the rotation angle applied by $G$. The state evolution at the $k$-th step then becomes:
\[
G^k\ket{\psi_0} = \cos(\frac{2k+1}{2}\cdot\tilde{\theta})\ket{\alpha} + \sin(\frac{2k+1}{2}\cdot\tilde{\theta})\ket{\beta} .
\]
The barrier function \eqref{grover_template} must satisfy the following modified constraint:
\[
B(c, G(z(\phi)) - B(c, z(\phi)) = c\cdot(\phi + \widetilde{\theta}) - c\cdot\phi \leq \delta
\]
which implies in the worst scenario of deviation ($\mu = \eta$):
\[
c \cdot (\tilde{\theta}) = c \cdot (\theta + \eta) \leq \delta.
\]
As per Algorithm \ref{algo},  we sample angles in the above defined sets $\mathcal{Z}_0, \mathcal{Z}_u$, and $\mathcal{Z} = \left\{ z(\phi) \mid \phi \in \left[0, 2\pi \right] \right\}$, and the constraints become linear with respect to the variables $\gamma,\space\lambda$ and $\delta$ (see Definition \ref{def: finiteHor}). We then solve the resulting linear optimization problem \eqref{eq:sb-BCoptprob} to find the values of the variables that satisfy the constraints. The formal verification of the constraints with the obtained values of $\gamma, \lambda $, $\delta $ and template coefficient $c$ (see \eqref{grover_template}) is then performed by the SMT solver. 

As we can see in Table \ref{tab: grover_example}, in all three Grover cases, a barrier certificate was found to guarantee the system's safety. For example, with parameters $\#Qubits$ = 5, $K = 32$, $M = 8$, and perturbed by $err$ = 0.5 and $\eta$ = 0.3, it was still possible to ensure Grover's safety with the following barrier certificate and values obtained through Algorithm \ref{algo}:
\[
B(z(\phi)) = 21.22065 \cdot \phi,
\]
with constraint parameters $\gamma = 11.49015$, $\lambda = 28.58842$, $\delta= 100$ and time horizon $T = 2$.


\subsection{Other Experiments}
We list here the sets of initial states and unsafe states (for $n$-qubit quantum systems) for each experiment presented in Table \ref{tab: infinite horizons examples} and Table \ref{tab: finite horizons examples}.
\begin{itemize}
    \item Z Gate:
    \begin{equation*}
        \begin{array}{l}
            \mathcal{Z}_0 = \{z \in \mathbb{C}^{2^n}: |z_0|^2 \geq 0.9\}, \text{ and}\\
             \mathcal{Z}_u = \{z \in \mathbb{C}^{2^n}: |z_0|^2 \leq 0.1\} 
        \end{array}
    \end{equation*}
    \item Controlled-NOT:
    \begin{equation*}
        \begin{array}{l}
            \mathcal{Z}_0 = \{z \in \mathbb{C}^{2^n}: |z_0|^2 \geq 0.9\}, \text{ and}\\
             \mathcal{Z}_u = \{z \in \mathbb{C}^{2^n}: \sum_{0 < i < 2^n} |z_i|^2\geq 0.11\} 
        \end{array}
    \end{equation*}
    \item T gate:
    \begin{equation*}
        \begin{array}{l}
            \mathcal{Z}_0 = \{z \in \mathbb{C}^{2^n}: |z_0|^2 \geq 0.9\}, \text{ and}\\
             \mathcal{Z}_u = \{z \in \mathbb{C}^{2^n}: |z_0|^2 \leq 0.1\} 
        \end{array}
    \end{equation*}
    \item Controlled-Z (CZ):
    \begin{equation*}
        \begin{array}{l}
            \mathcal{Z}_0 = \{z \in \mathbb{C}^{2^n}: |z_2|^2 + |z_3|^2 \geq 0.9\}, \text{ and}\\
             \mathcal{Z}_u = \{z \in \mathbb{C}^{2^n}: |z_2|^2 + |z_3|^2  \leq 0.05\} 
        \end{array}
    \end{equation*}
    \item Hadamard:
    \begin{equation*}
        \begin{array}{l}
            \mathcal{Z}_0 = \{z \in \mathbb{C}^{2^n}: |z_0|^2 \geq 0.9\}, \text{ and}\\
             \mathcal{Z}_u = \{z \in \mathbb{C}^{2^n}: |z_0|^2\leq 0.1\} 
        \end{array}
    \end{equation*}
    \item X Gate:
    \begin{equation*}
        \begin{array}{l}
            \mathcal{Z}_0 = \{z \in \mathbb{C}^{2^n}: \frac{1}{2^n} - err \leq |z_0|^2 \leq \frac{1}{2^n} + err\}, \text{ and}\\
             \mathcal{Z}_u = \{z \in \mathbb{C}^{2^n}: |z_1|^2 \geq 0.8\} 
        \end{array}
    \end{equation*}
    where $err = \frac{1}{10^{n+1}}$
    \item SWAP Gate:
    \begin{equation*}
        \begin{array}{l}
            \mathcal{Z}_0 = \{z \in \mathbb{C}^{2^n}: |z_1|^2 \geq 0.9\}, \text{ and}\\
             \mathcal{Z}_u = \{z \in \mathbb{C}^{2^n}: |z_0|^2\geq 0.5\} 
        \end{array}
    \end{equation*}
    \item Alternating CX and CZ: see Subsection \ref{example: cx cz}
    \item CX $\cdot $ CZ: same sets as the previous experiment, but with the following dynamics:
    \begin{equation*}
        f(z) = (CZ\cdot CX) z
    \end{equation*}
    \item Grover (Table \ref{tab: infinite horizons examples}):
    \begin{equation*}
        \begin{array}{l}
        \mathcal{Z}_0 = \{z \in \mathcal{Z}: \frac{1}{2^n} - \textit{err} \leq |z_j|^2 \leq \frac{1}{2^n} + \textit{err}, \\ -\sqrt{\text{err}} \leq \text{Im}(z_j) \leq \sqrt{\text{err}} \text{ per } 0 \leq j \leq 2^n - 1\} \\

            \mathcal{Z}_u^p = \{z \in \mathcal{Z}: |z_p|^2 >= 0.9\}
        \end{array}
    \end{equation*}
    where $err = \frac{1}{10^{n+1}}$ and $p$ is a non-solution state. 
    \item Grover* (Table \ref{tab: grover_example}): see Subsection \ref{Grover}.
    
\end{itemize}
The experiments: Z Gate, T Gate, alternation of CX and CZ and CX$\cdot$CZ in Table \ref{tab: infinite horizons examples} were generated using the barrier certificate of Theorem \ref{thm:k-inductive}, while the remaining ones were generated using Definition \ref{def:k-ind}. In Table \ref{tab: finite horizons examples} we instead show the results obtained using the barrier certificate of Definition \ref{def: finiteHor}. The averages of the generation and verification times were calculated over 10 independent repetitions of Algorithm \ref{algo}.

\section{Conclusions}
In this paper, we introduced a scenario-based methodology for the formal verification of quantum circuit using barrier certificates. We leveraged sampling-based optimization to efficiently synthesize certificates that prove a circuit will never reach undesired states, even over unbounded (infinite-horizon) executions or a finite number of steps. We successfully demonstrated our technique on various quantum circuits, and our results suggest that the scenario-based approach can model uncertainties, while maintaining computational feasibility.

Limitations of the proposed framework should be acknowledged. First, the approach currently relies on choosing a suitable function template (\eg polynomials of a given degree). Second, scalability remains a challenge: the number of samples needed grows with the state dimension (exponential in the number of qubits), and although we mitigated this with linear programming by sampling a finite set of states, extremely large quantum systems might still be computationally intensive to verify. Additionally, although the scenario approach offers high confidence and a pathway to a solution, if no certificate is found, one cannot be certain whether none exists for the given system.

Despite these limitations, this work opens several promising directions for future research. An immediate next step is to explore richer function classes for barrier certificates (\eg incorporating neural networks\cite{Fossil2,abate2024safe} or other nonlinear function families), while still retaining the ability to perform efficient optimization and SMT verification. Improving scalability is another key goal: for instance, more advanced sampling strategies or problem decomposition techniques could be employed to handle higher-dimensional state spaces and complicate constraints more efficiently (\eg generating uniform samples within semi-algebraic sets \cite{SemiAlgSampling} or computing certificates via decomposition \cite{nejati2022compositional}).

The certificates studied in this paper give only sufficient conditions for satisfying the requirements, rely on exact models of the quantum circuit, and do not take into account the statistics of the errors. In the future, we plan to utilize certificates that are necessary and sufficient \cite{majumdar2024necessary}, to develop techniques that do not require exact models of the underlying circuits \cite{schon2024data}, and to handle larger classes of specifications \cite{kordabad2024control}.

In conclusion, the paper presents a practical approach to quantum circuit verification by synthesizing barrier certificates via scenario-based optimization. The combination of theoretical rigor (formal certificates and SMT proof) and computational techniques (sampling and linear programming) provides a promising tool for ensuring quantum circuits behave as expected. We hope that this work will inspire further research into scalable formal verification for quantum systems, ultimately contributing to the reliability of future quantum computing systems and applications.

\section*{Acknowledgments}
The research of S. Soudjani is supported by the following grants: EIC 101070802 and ERC 101089047.
The research of P. Zuliani is supported by the SERICS project (PE00000014) under the Italian MUR National Recovery and Resilience Plan funded by the European Union - NextGenerationEU.






\bibliographystyle{IEEEtran.bst}
\bibliography{biblio}

\end{document}